\documentstyle[prd,aps]{revtex}
\begin{document}
\draft
\twocolumn[\hsize\textwidth\columnwidth\hsize\csname
@twocolumnfalse\endcsname
\preprint{ UMD-PP-97-22,
{}}
\title{ Anomalous $U(1)$ Mediated SUSY Breaking, Fermion Masses \\
and Natural Suppression of FCNC and CP Violating Effects}
\author{R.N. Mohapatra$^{(1)}$  and Antonio
Riotto$^{(2)}$}
\address{$^{(1)}${\it Department of Physics, University of Maryland,
College Park, Maryland 20742}}
\address{$^{(2)}${\it  Fermilab
National Accelerator Laboratory, Batavia, Illinois~~60510-0500}}
\date{August 26, 1996}
\maketitle
\begin{abstract}

We construct realistic supergravity models where supersymmetry 
breaking arises from the $D$-terms of an anomalous $U(1)$ gauge symmetry
broken at the Planck scale. The model has the attractive feature
that the gaugino masses, the $A$ and $B$ terms and
the mass splittings between the like-charged 
squarks of the first two generations compared to their average 
masses are all suppressed. As a result, the
electric dipole moment of the neutron  as well as the flavor changing
neutral current effects  are predicted to be naturally small.
These models predict naturally the expected value of the $\mu$-term and 
also have the potential to qualitatively explain the
observed mass hierarchy among quarks and leptons.

\end{abstract}
\pacs{
UMD-PP-97-22 
 }
\vskip2pc]

Supersymmetric extensions of the standard model (MSSM) have been the focus
of intense theoretical activity due to the fact that they provide a natural
solution to the problem of stability of the weak scale under quantum
corrections\cite{nilles}. 
Since experimental observations require supersymmetry to
be broken, it is essential to have a knowledge of
the nature and the scale of supersymmetry breaking in order to have a
complete understanding of the physical implications of these models.
At the moment, we lack such an understanding and therefore it is important 
to analyse the various ways that supersymmetry breaking can arise and
study their consequences, in the hope that one can gain some insight into
this problem. There are however several hints from the study of general
class of MSSM which could perhaps be useful in such a discussion. 
Two particular ones that rely on the supersymmetric sector of model are: 
(i) natural suppression of flavor changing neutral
currents (FCNC) which require a high degree of degeneracy among squarks
of different flavor and (ii) stringent upper limits on the electric
dipole moment of the neutron (NEDM) which imply constraints on the
gaugino masses as well as on the $A$ and $B$ terms of MSSM\cite{dine}. In this
letter we take the point of view that the above conclusions may be telling
us something about the nature of supersymmetry breaking. If this is true,
then it is important to isolate those SUSY breaking scenarios which 
realize the above properties in a simple manner and study their implications.
It has been already pointed out that the recently discussed gauge
mediated SUSY breaking models\cite{dine2} seem to have these
properties. In this letter, we study another class of models with
the same property and analyse its consequences. 

An important ingredient of the
 models, we are interested in, is the existence of an anomalous $U(1)$ gauge
symmetry whose linear $D$-term combined with an appropriate superpotential
for the hidden sector fields leads to supersymmetry 
breaking. This SUSY breaking
is fed down to the visible sector \cite{pomarol} both by the $D$-term as well as
by the supergravity effects. It was shown by Dvali and Pomarol in Ref.\cite{pomarol} that in
the resulting theory, the gaugino masses are suppressed. It was also
conjectured 
 that the FCNC and CP violating effects in these models are
suppressed. In this paper,
we construct full realistic versions of this model, which have
the feature that relative squark mass  difference (between the like-charged 
squarks of the first two generations) $\delta_{q}\equiv
\Delta m^2_{\tilde{q}}/m^2_{\tilde{q}}$, the gaugino masses
relative to the average squark masses $\delta_{\lambda}\equiv m_{\lambda}/
m_{\tilde{q}}$ as well as $\mu/m_{\tilde{q}}$ and $A/m_{\tilde{q}}$
are all small, with the suppression characterized by a  common 
parameter $\epsilon\simeq 10^{-2}$. This leads to the desirable property that
FCNC effects and SUSY CP effects in the 
electric dipole moment, $d^e_n$ of the neutron are
suppressed to an acceptable level. The suppression of $d^e_n$ is also
due to the factor $\epsilon$ unlike in Ref.\cite{pomarol}. Keeping
the above properties, we construct two models, which differ in the
way the electroweak symmetry breaking arises and the qualitative pattern
of fermion masses is predicted. In the first model, the electroweak
symmetry breaking arises at the tree level whereas in the second one,
it arises purely out of radiative corrections as in the usual supergravity
models. Furthermore the first model has the property th the down quark 
and charged
lepton masses are much smaller than the up quark masses of the corresponding
generation whereas
in the second one, the quark mass hierarchy is in more detailed
qualitative agreement with observations.
 Let us briefly outline the first model before proceeding to extract
its implications for the MSSM and illustrate how the afore-mentioned 
properties common to both the models emerge. At the end, we discuss the
second model, which shares all the properties with the first model
except the prediction for the fermion mass hierarchies and the way
the electroweak symmetry breaking is induced.
 
As already alluded to, the crucial feature of the model is the
existence of a $U(1)$ gauge group, which is anomalous.
The $U(1)$ group may be assumed to emerge from string theories.
We will assume that the anomaly is cancelled by the Green-Schwarz
mechanism. Since the $U(1)$ is anomalous, i.e. ${\rm Tr}{\bf Q}\neq 0$, a
Fayet-Illiopoulos term which is a linear  $D$-term is always generated 
as a quantum effect. We further assume that  there is a pair 
of hidden sector fields denoted by $\phi_+$ and $\phi_-$ 
which have $U(1)$ charges
$+1$ and $-1$ respectively and that the fields of the standard model
also carry $U(1)$ charges. It is the assignment of the $U(1)$ charges
to quark superfields that help in the solution of the FCNC and CP problems and in qualitatively explaining the fermion mass hierarchy.
We will illustrate the technique with the help of two models. In the first
one, the left-handed quark and
lepton doublets $Q$, $L$ are all assumed to have the $U(1)$ charge $q$ 
and the singlet fields $u^c$, $d^c$ and $e^c$ have charge $q'$;  
the two Higgs fields of MSSM,  $H_u$ and $H_d$ are assumed to have  
$U(1)$ charges $-q-q'$ and $1+q+q'$
respectively.  We will show that demanding that the superpotential
lead to $QH_dd^c$ type terms fixes the value of $q+q'$.  Note that both the
superpotential $W$ and the Kahler potential $K$ of the model 
must be invariant under the anomalous $U(1)$ symmetry. The superpotential is $W= W_0 + W_1 +W_2$,
where
\begin{eqnarray}
W_0&=&m\phi_+\phi_-,   \nonumber\\
W_1&=&h_u QH_u u^c,   \nonumber\\
W_2 &=& (h_d QH_d d^c+ h_e LH_de^c)\frac{\phi^2_-}{M^2_{P\ell}}\nonumber\\
&+&QH_u u^c\frac{\phi_+\phi_-}{M^2_{P\ell}}
+\cdot\cdot\cdot.
\end{eqnarray}	
In the above equation, the ellipses denote all other higher dimensional
terms allowed by the gauge symmetry and, as we will see below, make very
small contributions to the effects isolated below. 
The first term in $W_2$ fixes
$q+q'=1/2$. In what follows we will consider the assignment where
$q=q'=1/4$.
The parameter $m$ is chosen to be of the order of the weak scale.

Let us now write down the Kahler potential $K(z_i,z_i^*)$ for 
the fields of the  model generically indicated by $z_i$. It  can
be written as the sum of two terms: one that involves the bilinear
terms of the form $z_i^*z_i$ and a second piece that involves mixed
terms which are strongly constrained by the $U(1)$ symmetry.
\begin{eqnarray}
K&=&K_0 +K_1, \nonumber\\
K_0&=&\sum_i |z_i|^2,\nonumber\\
K_1&=& \lambda H_uH_d\frac{\phi^{\dagger}_+}{M_{P\ell}}+ {\rm h.c.} +\cdot\cdot\cdot.
\end{eqnarray}
In order to proceed further, we have to write down the potential
of the model involving the scalar fields $\phi_{\pm},~H^0_u,~H^0_d$ 
and isolate the vacuum state. 
The part of the potential containing the $\phi_{-}$ and $\phi_{+}$ fields reads
\begin{eqnarray}
V &=& m^2(|\phi_+|^2+|\phi_-|^2)\nonumber\\
&+&\frac{g^2}{2}
\left(-\frac{1}{2}|H^0_u|^2+\frac{3}{2}|H^0_d|^2+|\phi_+|^2-|\phi_-|^2+
\xi\right)^2.
\end{eqnarray}
Before discussing the minimization of the full potential, let us consider the
part of $V$ setting $H^0_u=H^0_d=0$. It is easy to see that its minimum
breaks supersymmetry as well as the anomalous $U(1)$ gauge symmetry 
with \cite{pomarol}
\begin{eqnarray}
\langle \phi_-\rangle &=&\left(\xi-{{m^2}\over{g^2}}\right)^{1/2},\:\:\langle \phi_+\rangle = 0\\ \nonumber
\langle F_{\phi_+} \rangle &=&m\left(\xi-{{m^2}\over{g^2}}\right)^{1/2}.
\end{eqnarray}
If we parameterize $\xi=\epsilon M^2_{P\ell}$, for $m\ll M_{P\ell}$, we have
$\langle \phi_-\rangle\simeq \epsilon^{1/2}M_{P\ell}$ and $\langle F_{\phi_+}
\rangle \simeq \epsilon^{1/2}m M_{P\ell}$. Assuming that $\xi$-term is
induced by loop effects, one can estimate\cite{pomarol,witten} 
$\xi={{g^2{\rm Tr}{\bf Q}M^2_{P\ell}}\over{192\pi^2}}$, so that $\epsilon$ 
can be assumed to be of order $10^{-2}$. It was pointed out in 
ref.\cite{pomarol} that the gaugino masses are generated in this model
by superpotential terms of type 
$ \lambda'W^{\alpha}W_{\alpha}\left(\frac{\phi_+\phi_-}{M^2_{P\ell}}\right)$.
As a result, one gets gaugino masses to be 
$m_{\lambda_{g}}=\lambda'\epsilon m$.

From the $K_1$ term in the Kahler potential supergravity effects  induce a
$\mu$-term  by means of  the Giudice-Masiero mechanism \cite{giudice}. 
Indeed, $K_1$ induces at low energy  the operator
\begin{equation}
\lambda\int d^4\theta  H_uH_d\frac{\phi^{\dagger}_+}{M_{P\ell}},
\end{equation}
giving rise to a $\mu$-term, with  $\mu=\lambda \epsilon^{1/2} m$. Notice 
that the corresponding $B$-term in the potential is not induced at order 
$\epsilon$, even though it will be generated by radiative corrections when 
running from the Planck scale down to the weak scale.

We integrate out the heavy field $\phi_{-}$ to obtain the 
effective potential of the light fields. Minimization with respect 
to  $\phi_{-}$ gives
\begin{equation}
|\phi_{-}|^2=\xi + |\phi_{+}|^2-\frac{1}{2}|H^0_u|^2+\frac{3}{2}|H^0_d|^2-\frac{m^2}{g^2}.
\end{equation}
The effective potential of the fields $(\phi_{+},H^0_d,H^0_u)$ is 
at the leading order in $m^2/M^2_{P\ell}$
\begin{eqnarray}
V&=&2 m^2 |\phi_{+}|^2+m^2_{H_u}|H_u^0|^2+m^2_{H_d}|H_d^0|^2\nonumber\\
&-&m_3^2\left(H_u^0H_d^0+\:\:{\rm h.c.}\right)
+\:D{\rm -terms},\nonumber\\
m^2_{H_d}&=&|\mu|^2+\frac{3}{2}m^2+m_0^2,\nonumber\\
m^2_{H_u}&=&|\mu|^2-\frac{1}{2}m^2+m_0^2,\nonumber\\
m^2_3&=&B\mu.
\end{eqnarray}
where we have indicated by "$D$-terms" the usual $D$-terms coming 
from $SU(2)\otimes U(1)$ and $m_0^2$ denotes the supersymmetry 
soft-breaking terms coming from supergravity, $m_0^2\sim \epsilon m^2$.  
A novel feature of this model is that 
the field $H^0_u$ gets  vacuum expectation value (VEV) already at the  
tree level since $m_{H_u}^2$ is negative at high scales. Since $B$ is 
not generated at order $\epsilon$,  to get the correct value of $M_Z^2$
 at the weak scale     
requires $m$ of the order of a few hundred GeV or less. It is then clear
that there is a potential conflict between the desirable value 
of $\langle H^0_u\rangle$
and the above prediction for the gaugino mass unless we 
choose a sufficient large coupling
 $\lambda'$. Furthermore, we do expect the 
renormalization group equations to reduce the $m^2_{H_u}$ as we go down to
the weak scale from the Planck scale. In any case, this model would lean more
towards a larger ${\rm tan} \beta$ sector of the MSSM. In the second model
that we present, the VEV of $H^0_u$ arises purely from radiative corrections
due to its different $U(1)$ charge assignment and no such constraint on
$\tan \beta$ or $\lambda'$ follow. We also notice that a VEV of order of 
$\frac{v_d v_u}{M_{P\ell}}$ is induced for the field $\phi_{+}$ when 
taking into account supergravity effects. 

Let us now look at other parameters of the theory. It is clear from the
the Eq. (1) that $A_u\sim \epsilon m$ whereas $A_d=0$ to order $\epsilon$.
In fact including the higher order terms in the superpotential and the Kahler
potential it is easy to see that $A_d\sim \epsilon^2 m$ (while 
$B\mu\sim \epsilon^{3/2} m^2$). Note however that these are the values at the
Planck scale and they will evolve to higher values at the weak scale. It is
however important to note that both the values of $A$ and $B$ remain of
order $\epsilon$ at most since the value of $B$ at weak scale is proportional
to $ m_{\lambda_g}$ times the renormalization logarithm factor and similarly
for $A$. Finally we note that the second term in the superpotential $W_2$
is the one responsible for the down quark and charged lepton masses. 
Substituting the VEV's for the $\phi_-$ field, it is easy to see that
there is an automatic suppression of $\epsilon$ in the down quark 
and charged lepton Yukawa couplings. If one chooses $h_{d,e}$ of the
same order as the up quark couplings, then this will explain why 
$m_{d_i,e_i}\ll m_{u_i}$, a property shared by the second and the
third generation fermions.

\noindent{\it Flavor Changing Neutral Current Effects}

Let us now discuss the FCNC effects in this model. To study this, we note
that squark masses $m^2_{\tilde{q}}$ (both left and right handed types) 
receive two
contributions: a universal contribution from the $D$-term which is of 
order $m^2$ and a non-universal contribution from the supergravity Kahler
potential of order $ F^2_{\phi_+}/M^2_{P\ell}\equiv \epsilon m^2$. 
As both these 
contributions are extrapolated from the Planck scale down to the weak scale
the pattern of the first two generation squark masses remain practically
unchanged whereas the masses of the stop receive significant contributions.
It was noted in \cite{dine} that in order to satisfy the present observations
of FCNC effects (such as $K^0-\bar{K}^0$ mixing), the mixings between the
$\tilde{s}$ and the $\tilde{d}$ squarks (i.e. $m^2_{\tilde{s}\tilde{d}}$)
in the flavor basis or the squark mass differences 
between the first two generations in the mass basis
must satisfy a stringent constraint. In the flavor basis, it is given by
(see Dugan et al., in \cite{dine}), Im$\left({{m^4_{\tilde{s}\tilde{d}}}\over
{m^4_{\tilde{q}}}}\right)\leq 6\times 10^{-8}{{m^2_{\tilde{q}}}\over{m^2_W}}$.
We have assumed the phases in our model to be arbitrary; therefore the most
stringent constraint comes from the CP-violating part of the $K^0-\bar{K}^0$
mass matrix.
In our model, $m^2_{\tilde{s}\tilde{d}}$ arises purely from the supergravity
effects are of order $\sim \epsilon m^2$ and the above FCNC constraint
is satisfied if $\epsilon\simeq 10^{-2}$ or so. Thus our model
confirms the conjecture of Ref.\cite{pomarol}.

\noindent{\it Electric dipole moment of the neutron}

The electric dipole moment of the neutron $d^e_n$ 
in supersymmetric  models have been
discussed in several papers\cite{nedm} and it is by now well-known that
the gluino intermediate states in the loop graph contributing to the
$d^e_n$ gives a contribution which is some three orders of magnitude
larger than the present experimental upper limit
for generic values of the parameters. The situation is different in our
model since we see that a number of parameters of the model such as the
gluino masses, the $A$ and $B$ are down by powers of $\epsilon$. In order
to see the impact of this on the NEDM, we will again consider the charge
assignment for the first model 
where the Kahler potential induced mass splittings
in the squark masses are of order $\epsilon m^2$. For the gluino contribution,
we borrow from the calculation of Kizukuri and Oshimo \cite{nedm},
which gives:
\begin{eqnarray}
d^e_n&=& \frac{2e\alpha_s}{3\pi}\left( \sin\alpha_u A_u -
\sin\theta_\mu {\rm cot}\beta |\mu|\right)\nonumber\\
&\times&\frac{m_u}{m^2_{\tilde{q}}}\frac{1}{
m_{\lambda_3}}I\left(\frac{m^2_{\tilde{q}}}{m^2_{\lambda_3}}\right),
\end{eqnarray}
where $\alpha_u=\theta_{A_u}-\theta_{\lambda_3}$ is the differerence 
between the phases of the $A$-term and the gluino mass. 
We have kept only the up quark contribution since in our
model $A_d \ll A_u$; $m_{\tilde{q}}$ denotes the mass of the heavier of
the two eigenstates. Since in this model, $m_{\lambda_3}\simeq \sqrt{\epsilon}
m$ and $m_{\tilde{q}}\simeq m$, one 
finds that $I\simeq \epsilon$. This  leads to  
$d^e_n\simeq {{2\alpha_s}\over{3\pi}}\epsilon^{3/2} {{m_u}\over
{m^2}}$. Here we have used the fact that $A\sim \epsilon m$; 
$\mu \sim \sqrt{\epsilon}m$. For $\epsilon \simeq 10^{-2}$, this 
gives an additional suppression of $10^{-3}$ over the prediction of generic
parameter values of the MSSM as required.   

We wish to point out that the above suppression depends on the fact
that $Q_1,u^c_1,d^c_1$ all have nonzero $U(1)$ charge. If on the other hand,
$d^c$ and $u^c$ had zero charge, their dominant mass would come
from the supergravity effect and, as a result,  
$m^2_{\tilde{d^c}}\sim m^2_{\tilde{u^c}}\simeq \epsilon m^2$. 
The the above gluino 
contribution to $d^e_n$ would then be less suppressed (by a factor
$\sqrt{\epsilon}$ rather than $\epsilon^{3/2}$). 

\noindent{\it A second model:}

We next present an alternative charge assignment which qualitatively explains
the observed
mass hierarchy of quarks while keeping all other $\epsilon$-suppressions
of the various parameters of the
model unchanged. We choose $Q_3$, $u^c_3$ and $H_u$ 
to have zero $U(1)$ charge, but all other quarks have charge $+1$
as does $H_d$. This charge assignment for the $H_{u,d}$ allows the
Kahler potential term $K_1$ in Eq. (2) so that the suppression of the
$\mu$-term is maintained as in the first model.
Moreover, only the Yukawa coupling $Q_3H_uu^c_3$ is allowed
without any suppression from the $\epsilon$ factor explaining why the
top quark has large mass \cite{jain}. On the other hand, the other
Yukawa couplings are suppressed with powers of $\epsilon$ qualitatively
explaining why their masses are so much smaller than the top quark mass. The
superpotential for such a theory can be written as:
\begin{eqnarray}
& & h_{33} Q_3H_uu^c_3 + h'_{3a} Q_3H_dd^c_a\frac{\phi^2_-}{M_{P\ell}}
+h_{3j}Q_3H_uu^c_j\frac{\phi_-}{M_{P\ell}}\nonumber\\
&+&h_{ij}Q_iH_uu^c_j\frac{\phi^2_-}{M^2_{P\ell}}
+h'_{ij}Q_iH_dd^c_j\frac{\phi^3_-}{M_{P\ell}}
\end{eqnarray}
where $i,j$ go over 1, 2 and $a$ goes 
over 1, 2 and 3 and are generation indices.
They lead to the following kind of up and down quark mass matrices.
\begin{eqnarray}
M_u=m_1\left(
\begin{array}{ccc}
\epsilon & \epsilon & \sqrt{\epsilon} \\
\epsilon & \epsilon & \sqrt{\epsilon} \\
\sqrt{\epsilon} & \sqrt{\epsilon} & 1
\end{array}\right)
\end{eqnarray}
and
\begin{eqnarray}
M_d=m_2
\left(\begin{array}{ccc}
\epsilon^{3/2} &\epsilon^{3/2} & \epsilon \\
\epsilon^{3/2} & \epsilon^{3/2} & \epsilon \\
\epsilon & \epsilon & \epsilon
\end{array}\right).
\end{eqnarray}
where $m_{1,2}$ are mass parameters related to the $v_{u,d}$ and the
Yukawa couplings.
This pattern predicts that $m_u=0$ and $m_c\simeq \epsilon m_t$ . In
the down sector, $m_b\simeq \epsilon m_t$, $m_s\simeq
\sqrt{\epsilon}m_b$ and $m_d=0$, 
which for $\epsilon\simeq 10^{-2}$ roughly corresponds
to observations. Similar considerations can be applied to the leptons.
Other than to note this qualitatively interesting prediction,
we do not want to pursue the detailed predictions of this model
for fermion masses and mixings here. However, we want to
point out an interesting feature of the model that $H_u$ VEV arises purely
from radiative corrections (and not at the tree level as the first class
of models) and is therefore not locked to the value of $m$. As far as the
$A$-term is concerned, it is clearly suppressed by powers of $\epsilon$
which depend on then generation indices; for instance, $A^u_{33}=\epsilon m$,
$A^d_{3a}=\epsilon^{3/2} m$ and so on. In this case the $d^e_n$ is more highly
suppressed than the first model.

A few comments are in order regarding various aspects of the model:

\noindent(i) The low energy effective theory contains all fields except
the $\phi_-$ and all components of the field $\phi_+$ have masses of order
$m$ and they do not mix with the other Higgs fields even though the U(1)
symmetry is broken.

\noindent (ii) Different versions of our theory with other charge assignments
are possible. But one has to be careful not to assign negative $U(1)$ charges
to the quarks or leptons since that will lead to breaking of color and 
electric charge.

\noindent (iii) The gravitino mass in this model is of the order of
$\sqrt{\epsilon}v_uv_d m/M^2_{P\ell}$ or less and it arises 
once the $\phi_+$ field acquires a VEV due to supergravity effects.

\noindent (iv) The model has the feature that one can choose the Kahler
potential and the superpotential
with arbitrary number of higher dimensional terms as long they
are $U(1)$ invariant and yet our results will remain unchanged. The
higher order terms induce small corrections down by higher powers 
in $\epsilon$.

In conclusion, we have demonstrated with two examples that
it is possible to construct interesting realistic supersymmetric models 
of quarks and leptons using
the idea that an anomalous $U(1)$ gauge symmetry is responsible for
generating supersymmetry breakdown. These models have the additional
attractive feature that they solve
several fine tuning problems of the MSSM associated with FCNC effects and
electric dipole moment of the neutron. They also give desirable values for
the $A$, the $B$ and the $\mu$ parameters 
and also have the potential to qualitatively
explain fermion mass hierarchies.

\vskip 0.1 cm

The work of R. N. M. is supported by the NSF grant no. PHY-9421385
and the work of A. R. is supported by the DOE and NASA under Grant NAG5--2788.

\end{document}